\documentclass[trackchanges]{aastex701}

\usepackage{graphicx}
\usepackage{dcolumn}
\usepackage[table]{xcolor} 
\usepackage{colortbl}
\usepackage{bm}
\usepackage{comment}
\usepackage{xcolor}
\usepackage{hyperref}
\hypersetup{
    colorlinks=true,
    linkcolor = blue,
    urlcolor = magenta,
    citecolor = blue,
}
\usepackage{url}

\usepackage{booktabs}
\usepackage{array}
\usepackage{amsmath}

\begin{document}

\title{Velocity space origins of pressure-strain interaction in multi-population distributions and its application to magnetic reconnection}

\author{M.~Hasan Barbhuiya}
\affiliation{Department of Physics and Astronomy, 
Clemson University, Clemson, SC 29634, USA}
\altaffiliation{Department of Physics and Astronomy and the Center for KINETIC Plasma Physics,  West Virginia University, Morgantown, WV 26506, USA}
\email{mbarbhu@clemson.edu}

\author{Paul A.~Cassak}
\affiliation{Department of Physics and Astronomy, 
Clemson University, Clemson, SC 29634, USA}
\altaffiliation{Department of Physics and Astronomy and the Center for KINETIC Plasma Physics,  West Virginia University, Morgantown, WV 26506, USA}
\email{pcassak@clemson.edu}

\author{Sarah Conley}
\affiliation{Department of Physics and Astronomy, Bates College, Lewiston, ME 04240, USA}
\email{sconley2@bates.edu}

\author{Julia E. Stawarz}
\affiliation{School of Engineering, Physics, and Mathematics, Northumbria University, Newcastle upon Tyne, NE1 8ST, UK}
\email{julia.stawarz@northumbria.ac.uk}

\author{Emily Lichko}
\affiliation{Plasma Physics Division, U. S. Naval Research Laboratory, Washington, DC, 20375, USA}
\email{emily.r.lichko.civ@us.navy.mil}

\author{Jason TenBarge}
\affiliation{Department of Astrophysical Sciences, Princeton University, Princeton, New Jersey 08544, USA}
\email{tenbarge@princeton.edu}

\author{James Juno}
\affiliation{Princeton Plasma Physics Laboratory, Princeton, NJ, 08540, USA}
\email{jjuno@pppl.gov}

\author{Jason R. Shuster}
\affiliation{Space Science Center, University of New Hampshire, Durham, NH, USA}
\email{Jason.Shuster@unh.edu}

\author{Gregory G. Howes}
\affiliation{Department of Physics and Astronomy, University of Iowa, Iowa City, IA 52245, USA}
\email{gregory-howes@uiowa.edu}

\author{Subash Adhikari}
\affiliation{Department of Physics and Astronomy, University of Delaware, Newark, DE 19711}
\email{subash@udel.edu}

\begin{abstract}
A forefront research question is how energy evolves in weakly collisional plasmas for which departures from local thermodynamic equilibrium (LTE) are significant. The standard approach is studying the terms in the non-LTE energy evolution equation derived by taking the second moment of the Boltzmann equation, but the resultant fluid metrics do not retain information about which particles at which velocities drive energy evolution. A widely studied channel for internal energy density evolution is the pressure-strain interaction. Here we employ the kinetic pressure-strain [S. A. Conley et al., {\it Phys. Plasmas,} {\bf 31}, 122117 (2024)], a phase space diagnostic whose velocity-space integral recovers the pressure-strain interaction to disambiguate the contributions to pressure-strain interaction from disparate particle populations in composite phase-space densities. We develop phase-space analogs of the pressure-strain interaction decompositions to provide the phase-space origins of normal vs. sheared flow. 
We introduce the ``kinetic strain-rate" tensor, the phase-space analog of strain-rate tensor, which we argue is needed to interpret phase-space origins of pressure-strain interaction. To demonstrate the utility of these quantities, we investigate them for composite electron distributions near the electron diffusion region in two-dimensional particle-in-cell simulations of antiparallel symmetric magnetic reconnection. We find that the phase space-based diagnostics isolate the roles of distinct populations. These results contribute to a growing body of work providing new methods for quantifying phase space energy evolution for a broad array of processes, from magnetic reconnection to collisionless shocks and turbulence, opening new pathways for answering longstanding problems of particle energization in weakly collisional plasmas.
\end{abstract}



\section{Introduction}
\label{sec:intro}

Classical plasma and fluid systems dominated by collisions generally approach local thermodynamic equilibrium (LTE), where the evolution of the internal energy is relatively well-understood.
However, for systems where collisions are weak, {\it i.e.,} when the collisional time scales are longer than the dynamical time scales of the system, the study of energy evolution is significantly more challenging.
Understanding the evolution of internal energy in weakly collisional plasmas is important because they are ubiquitous in the heliosphere, including Earth’s magnetosphere, the solar wind, planetary magnetospheres, and laboratory plasmas relevant to space conditions \citep{Burch16b,Howes17,Marsch2018solar,Shuster19,Wang_2019_GRL,Matthaeus20,Shi_2022_PRL,Ren_2024_GRL}, where non-LTE processes generate non-Maxwellian phase-space densities. 

The standard approach to analyze the evolution of the internal energy density has been to study the terms in the non-LTE energy density evolution equation derived by taking the second moment of the Boltzmann/Vlasov equation and isolating the evolution of the energy density contained in the peculiar/random motion \citep{Gartenhaus_1964}. Numerous terms describing different avenues for internal energy density evolution arise \citep{Gartenhaus_1964}. For this study, we focus on one called the pressure-strain interaction \citep{del_sarto_pressure_2016,Yang17,yang_PRE_2017,Yang_2022_ApJ,Cassak_PiD1_2022}.

It is a power density describing the time rate of change to the internal energy density due to the presence of non-uniform bulk flow. 
The system-integrated pressure-strain interaction is the only source/sink term for internal and bulk kinetic energies for closed, periodic, or isolated systems, such as simulations of decaying turbulence \citep{Yang_2022_ApJ}. The pressure-strain interaction has been studied in reconnection and turbulence simulations \citep{Yang_2022_ApJ, song_forcebalance_2020,Barbhuiya_PiD3_2022,Barbhuiya_PRE_2024, Barbhuiya_phases_2025}, and has also been demonstrated to be measurable using NASA's Magnetospheric Multiscale (MMS) mission \citep{zhou_measurements_2021, Chasapis18,bandyopadhyay_energy_2021, Burch_PoP_2023}. The pressure-strain interaction can be decomposed to isolate various physical effects. One decomposition is to split apart the effects of compressibility in pressure dilatation and incompressibility in so-called ${\rm Pi-D}$ \citep{Yang17,yang_PRE_2017}. (See Ref.~\citep{Adhikari_2025_Helmholtz_PoP} for a different assessment of this decomposition).
An alternate decomposition of the pressure-strain interaction is to split apart the effects of normal flow and sheared flow into so-called PDU and ${\rm Pi-D}_{\rm shear}$ (which will be defined analytically in Sec.~\ref{sec:kpsdecomposition}), respectively \citep{Cassak_PiD1_2022}.

While the pressure-strain interaction has led to significant strides in our understanding of the evolution of internal energy in weakly collisional systems, there is a notable drawback. The process of taking the second moment of the Boltzmann/Vlasov equation is carried out via an integral over all velocity space. 
Consequently, the pressure-strain interaction does not retain information about how particles at different velocities contribute differently to it. However, see Ref.~\citep{Cassak_PiD1_2022} that described the velocity-space mechanisms underlying the pressure-strain interaction, even though its kinetic information is integrated out.)
This loss of velocity-space information prevents one from understanding the underlying kinetic origin of the pressure-strain interaction.

An analogous situation arises for the energy conversion between electromagnetic fields and the charged particles in a plasma. From Poynting's theorem, the fluid form of the power density of this conversion is ${\bf J} \cdot {\bf E},$ where ${\bf J}$ is the current density and ${\bf E}$ is the electric field, and a positive ${\bf J} \cdot {\bf E}$ indicates energy transfer from the electric field to the charged particles. As with pressure-strain interaction, the ``fluid" form of the power density lacks information about the kinetic, \textit{i.e.,} phase space, physics that leads to there being a non-zero power density.

A key breakthrough occurred in the last decade \citep{Klein16}. 
The key to overcoming this loss of information is to analyze the particle energy density at the phase space level, {\it i.e.,} before integrating over velocity space. As we will review in Sec.~\ref{sec:review}, scaling the Boltzmann/Vlasov equation by $(1/2)m v^2$, where $m$ is the mass of a constituent particle and ${\bf v}$ is the velocity space coordinate, leaves an equation for the density of particle energy in phase space.  One term in this equation, related to what the authors called the ``field-particle correlation" (FPC), gives the conversion of phase-space energy density between the electric field and the charged particles as a function of the position and velocity phase space coordinates. The velocity space integral of the term gives $-{\bf J}_\sigma \cdot {\bf E}$, where ${\bf J}_\sigma$ is the current density solely due to species $\sigma$. The FPC is said to be the phase-space analog of ${\bf J}_{\sigma} \cdot {\bf E}$.

The FPC has emerged as a critical diagnostic and has been widely employed to identify mechanisms responsible at the phase space level for energy density conversion between the electromagnetic fields and the total plasma kinetic (bulk flow and internal) energy density.
Examples include ion Landau damping vs.~ion cyclotron damping \citep{Klein20,Afshari_NatCom_2024}, bulk electron acceleration in magnetic reconnection exhausts and near the X-line \citep{McCubbin22}, single vs.~multiple rounds of shock-drift acceleration of ions in one and three-dimensional collisionless shocks \citep{Juno_2023_ApJ}, and ion energization by instability-driven electric fields in a shock ripple \citep{Brown_2023_JPP}. We emphasize that the FPC has proven valuable in elucidating the role of energization of resonant particles in resonant wave-particle interactions, as well as in distinguishing between different particle populations, to determine which process drives the energy evolution in both simulations and observations of plasma systems \citep{Howes17,Klein17,Chen19,Afshari_JGR_2021,Klein20,Juno21,McCubbin22,Montag22,Montag2025ApJ}.

Recently, it was noted that a treatment analogous to the FPC could be carried out for the pressure-strain interaction \citep{Conley_2024_PoP}.  The result is the phase-space analog of the pressure-strain interaction, which is called the kinetic pressure-strain (KPS). It reveals the phase-space origin of the pressure-strain interaction, which is a key term in determining the evolution of internal energy density of a system as opposed to the evolution of the total of the kinetic and internal energy density that is determined using the FPC.
In Ref.~\citep{Conley_2024_PoP}, KPS was utilized to analyze resonant interactions in two different waves, where it was found that KPS elucidated the kinetic physics taking place during Landau damping, with resonant electrons playing a major role.

In the present study, we address several aspects of KPS.  First, we argue that KPS, similar to the FPC, should be valuable to study internal energy density evolution due to disparate populations of particles.
Second, we argue that a thorough understanding of the phase-space origins of the pressure-strain interaction must be accompanied by the phase-space origin of the strain-rate tensor. 
The strain-rate tensor, which appears in the pressure-strain interaction, is also a fluid quantity and thus retains no information about the phase-space structures giving rise to it.
We introduce the phase-space analog of the strain-rate tensor, which we call the kinetic strain-rate (KSR) tensor, to ascertain the ``kinetic" origins of strain-rate tensor elements.
Evaluating the KSR provides insight into where in phase space the particles are located that give rise to the dominant contribution to the strain-rate tensor. 
Interestingly, we find the particle population(s) contributing dominantly to the pressure-strain interaction need not be the same as the particle population(s) giving the dominant contributions to the strain-rate tensor. 
This distinction is especially important in multi-population phase-space densities: KSR identifies which population generates the local flow kinematics, whereas KPS identifies which population is responsible for the dominant local internal energy density evolution through pressure-strain interaction.
We show that near the reconnection site of symmetric antiparallel magnetic reconnection, these two roles can be carried by different electron populations.
Another thread of the present work is to investigate the phase-space analogs of the decomposition of pressure-strain interaction into PDU and ${\rm Pi-D}_{{\rm shear}}$.  First, we derive the evolution equation for the phase-space analog of the internal energy density directly from the Boltzmann equation in a stationary reference frame, unlike in Ref.~\citep{Conley_2024_PoP}, which was done in the plasma rest frame.  The kinetic pressure-strain appears straightforwardly in this derivation.
This analysis reveals the phase-space analogs for all of the terms that can change the internal energy density. Moreover, it is straightforward to obtain the phase-space analogs for the different decompositions that have been developed for the pressure-strain interaction.

We apply the theoretical framework discussed in the prior paragraph to a well-studied phenomenon that produces phase-space densities with multiple populations: collisionless anti-parallel magnetic reconnection.
Using two–dimensional particle-in-cell simulations, we examine how the phase-space analogs reveal the underlying velocity-space origins of the pressure-strain interaction and its decompositions. 
We find that (1) the kinetic pressure-strain successfully disambiguates the roles of different populations, and its decomposition into phase-space analogs of PDU and ${\rm Pi-D}_{\rm shear}$ reveals whether the effect is due to normal or sheared bulk flow, (2) the dominant population causing pressure-strain interaction can be a population with a relatively small phase-space density, (3) while the population that has the dominant contribution to the pressure-strain interaction may be the same population that gives the dominant contribution to the strain-rate tensor, there are also circumstances for which different populations can be dominant for the two different terms, and (4) the kinetic pressure-strain interaction can clarify when there are multiple populations and the contribution from two (or more) populations opposes each other at the same physical location. We argue that the framework studied in this paper will be useful for an array of weakly collisional plasma processes beyond resonant wave-particle interactions and reconnection, including turbulence and collisionless shocks.
The present study complements several recent numerical and observation studies that analyzed the velocity space structure of the terms in the Vlasov equation \citep{Shuster21,shuster_2023_JGR,Shuster2026_Comm_Phys} 
and the phase-space analog of the momentum equation \citep{shuster_electron-scale_2021_PoP,Norgren_2025_PoP,Shuster2026_Comm_Phys}.  
These studies made the connection between velocity-space structures of non-LTE phase-space densities and fluid moment equations.

The layout of this paper is as follows.  The theory is discussed in Sec.~\ref{sec:theory}. Section~\ref{sec:simulation} describes the numerical simulation carried out in this study.  The simulation results are discussed in Sec.~\ref{sec:results}. Section~\ref{sec:conclusion} summarizes the results and discusses further implications.

\section{Theory}
\label{sec:theory}

\subsection{Brief Review}
\label{sec:review}

The field-particle correlation (FPC) analysis is carried out 
starting from the Boltzmann transport equation for species $\sigma$ (where we suppress the species in the notation unless needed for clarity):
\begin{equation}
    \frac{\partial f}{\partial t} + \boldsymbol{\nabla} \cdot ({\bf v} f) +  \boldsymbol{\nabla}_v \cdot \left(\frac{{\bf F}}{m} f \right) = C[f],
    \label{eq:boltzeqn}
\end{equation}
where $f$ is the phase-space density of species $\sigma$, $t$ is time, ${\bf v}$ is the velocity space coordinate, $\boldsymbol{\nabla}$ and $\boldsymbol{\nabla}_v$ are the position- and velocity-space gradient operators, respectively, ${\bf F}$ is the net body force which for magnetized plasmas (neglecting other body forces) is $q({\bf E} + {\bf v}\times {\bf B}/c)$ (in cgs units), $q$ and $m$ are the charge and mass of the constituent of species $\sigma$, respectively, ${\bf B}$ is the magnetic field, $c$ is the speed of light, and $C[f]$ is the collision operator which we do not specify for generality, but we note that it
may be a sum of intra-species and inter-species collision operators given by $C[f] = C_{\sigma}[f_\sigma] + \sum_{\sigma^\prime \neq \sigma} C_{\sigma \sigma^\prime}[f_\sigma,f_{\sigma^\prime}]$.  Multiplying Eq.~\ref{eq:boltzeqn} by $(1/2)m v^2$, and manipulating the result gives the time evolution equation for the phase space energy density $w = (1/2) m v^2 f$
\citep{Klein16}, 
\begin{equation}
    \begin{split}
    \frac{\partial w}{\partial t} & + ({\bf v} \cdot \boldsymbol{\nabla}) w +  \frac{q v^2}{2} {\bf E} \cdot \boldsymbol{\nabla}_v f \\ & + \frac{q v^2}{2} \left(\frac{{\bf v}}{c} \times {\bf B} \right) \cdot \boldsymbol{\nabla}_v f = \frac{1}{2} m v^2 C[f].
    \label{eq:psweqn}
    \end{split}
\end{equation}
When combined, the third and fourth terms on the left-hand side of Eq.~\ref{eq:psweqn} describe the coupling of the electromagnetic force to the velocity-space variations in $f$, and we dub them the ``force-particle interaction" terms. The electric field term, \textit{i.e.,} the third term on the left-hand side, is used to construct the FPC \citep{Klein16}. 

The second term gives rise to the kinetic pressure-strain. 
We first write it as $\nabla \cdot ({\bf v}w)$. To separate the individual elements that contribute to the KPS, the substitution ${\bf v} = {\bf u} + {\bf v}^{\prime}$ is introduced,
where ${\bf u} =  (1/n)\int \mathbf{v} f d^3v$ is the bulk flow velocity, $n = \int f d^3v$ is the number density, and ${\bf v}^\prime$ is the peculiar/random velocity.  One of the numerous terms that results from this substitution is the so-called alternative KPS,
\begin{equation}
    \widetilde{K}_{PS} = -m v_j^\prime v_k^\prime f \partial_j u_k,
    \label{eq:altKPS}
\end{equation}
where $\partial_j$ is the $j$'th component of $\mathbf{\nabla}$ and we use the Einstein summation convention on repeated indices. 

\subsection{Evolution equation for internal energy density in phase space}

Our goal is to manipulate the non-relativistic Boltzmann transport equation to obtain an evolution equation for
the phase space internal energy density, which we define as $w_{\rm int} = (1/2) m v^{\prime 2} f$. 
Multiplying Eq.~(\ref{eq:boltzeqn}) by $(1/2) m v^{\prime 2}$ gives
\begin{equation}
    \frac{1}{2} m v^{\prime 2} \frac{\partial f}{\partial t}  + \frac{1}{2} m v^{\prime 2} \boldsymbol{\nabla} \cdot ({\bf v} f) +  \frac{1}{2} m v^{\prime 2} \boldsymbol{\nabla}_v \cdot \left(\frac{{\bf F}}{m} f \right) = \frac{1}{2} m v^{\prime 2} C[f].
    \label{EQ:PSINTEQ1}
\end{equation}

Using the product rule, the first term is equivalent to
\begin{equation}
     \frac{1}{2} m v^{\prime 2} \frac{\partial f}{\partial t} = \frac{\partial w_{{\rm int}}}{\partial t} 
     + m f {\bf v}^\prime \cdot \frac{\partial {\bf u}}{\partial t}, 
     \label{EQ:PSINTEQ1_1}
\end{equation}
where we use $\partial {\bf v}^\prime/\partial t = \partial ({\bf v}-{\bf u}) / \partial t = - \partial {\bf u} / \partial t$ in the last term.
Using an analogous manipulation on
the second term in Eq.~\ref{EQ:PSINTEQ1} gives
\begin{equation}
    \frac{1}{2} m v^{\prime 2} \boldsymbol{\nabla} \cdot ({\bf v} f)  = \boldsymbol{\nabla} \cdot \left( \frac{1}{2}m v^{\prime 2} {\bf v}^{\prime} f\right)
    + \boldsymbol{\nabla} \cdot \left( \frac{1}{2}m v^{\prime 2} {\bf u} f\right) + m v_j^{\prime} v_k^{\prime} f \partial_j u_k + m u_j v_k^{\prime} f \partial_j u_k,
    \label{EQ:PSINTEQ1_2}
\end{equation}
where the last two terms are written using the Einstein summation convention.
Collecting terms from Eq.~\ref{EQ:PSINTEQ1}-\ref{EQ:PSINTEQ1_2}, we obtain the time evolution equation for $w_{\rm int}$ in the stationary (laboratory) frame,
\begin{equation}
\frac{\partial w_{\rm int}}{\partial t} + \boldsymbol{\nabla} \cdot \left( w_{{\rm int}} {\bf u} \right) 
+ \boldsymbol{\nabla} \cdot \left( \frac{1}{2}m v^{\prime 2} {\bf v}^{\prime} f \right) + m v_j^{\prime} v_k^{\prime} f \partial_j u_k + m u_j v_k^{\prime} f \partial_j u_k + m f {\bf v}^\prime \cdot \frac{\partial {\bf u}}{\partial t} + \frac{1}{2} v^{\prime 2} \boldsymbol{\nabla}_v \cdot \left({\bf F} f \right) = \frac{1}{2} m v^{\prime 2} C[f]. 
\label{EQ:PSINTEQ2}
\end{equation}

To elucidate the physical meaning of the terms in this equation, we make connections between the phase space internal energy density evolution equation for $w_{{\rm int}}$ and the evolution equation for the (fluid) internal energy density $\mathcal{E}_{\rm int} = \int w_{{\rm int}} d^3v = (3/2) \mathcal{P} = (3/2) n k_B \mathcal{T}$, where 
$\mathcal{P}=(1/3) {\rm Tr}({\bf P})$ is the effective pressure,
${\bf P}$ is the pressure tensor with elements $P_{jk} = \int m v_{j}^\prime v_{k}^\prime f d^3v$, $\mathcal{T}$ is the effective temperature, and $k_B$ is Boltzmann's constant. This connection is achieved by integrating Eq.~\ref{EQ:PSINTEQ2} over all velocity space, which we treat term by term. 

The velocity space integral of the first term in Eq.~\ref{EQ:PSINTEQ2} directly gives $\partial \mathcal{E}_{\rm int}/\partial t$.  Similarly, 
the velocity space integral of the second term yields the divergence of the internal energy density flux 
$\nabla \cdot (\mathcal{E}_{\rm int} {\bf u})$. The velocity space integral of the third term is the divergence of the vector heat flux density $\nabla \cdot {\bf q}$, where ${\bf q} = \int (1/2) m v^{\prime 2} {\bf v}^\prime f d^3v$ is the vector heat flux density.
The velocity space integral of the fourth term is $({\bf P} \cdot \boldsymbol{\nabla}) \cdot {\bf u}$, the negative of the pressure-strain interaction. Thus, as was previously shown in Ref.~\citep{Conley_2024_PoP}, this term is the phase-space analog of pressure-strain interaction defined in Eq.~\ref{eq:altKPS}, 
so that $-({\bf P} \cdot \boldsymbol{\nabla}) \cdot {\bf u} = \int \widetilde{K}_{\rm PS} d^3v$.   
The fifth and sixth terms are linear in ${\bf v}^\prime$, so they both are identically zero when integrated over velocity-space; thus, neither has a fluid analog. 
For the last term on the left-hand side of Eq.~\ref{EQ:PSINTEQ2}, the force term, we demonstrate in Appendix \ref{appensec:A} that it vanishes identically when integrated over velocity-space, which reiterates that body forces cannot locally change the internal energy density of the plasma.  Lastly, the velocity space integral of the collision term on the right-hand side gives $\dot{Q}_{{\rm coll}} = \int (1/2) m v^{\prime 2} \sum_{\sigma^\prime \neq \sigma} C_{\sigma \sigma^\prime}[f_\sigma,f_{\sigma^\prime}] d^3v$, the volumetric heating rate due to inter-species collisions. The overall result for the fluid internal energy density evolution equation from the integral of Eq.~\ref{EQ:PSINTEQ2} has the well-known form
\begin{equation}
    \frac{\partial {\cal E}_{{\rm int}}}{\partial t} 
    + \boldsymbol{\nabla} \cdot 
    ({\cal E}_{{\rm int}} {\bf u} + {\bf q}) 
    = - ({\bf P} \cdot \boldsymbol{\nabla}) \cdot {\bf u} 
    + \dot{Q}_{{\rm coll}}.
\label{eq:IntEnerEq}
\end{equation}

In this study, we focus on the alternative kinetic pressure-strain and leave other terms for future work. 
As with its fluid analog, \textit{i.e.,} the pressure-strain interaction, the negative sign in the definition of $\widetilde{K}_{\rm PS}$ is included in Eq.~\ref{eq:altKPS}. Thus, the physical interpretation of $\widetilde{K}_{\rm PS}$ is that $\widetilde{K}_{\rm PS}>0$ in isolation contributes to an increase in $w_{\rm int}$, while $\widetilde{K}_{\rm PS}<0$ in isolation contributes to a decrease in $w_{\rm int}$. However, we emphasize that, as illustrated by Eq.~\ref{EQ:PSINTEQ2}, other quantities can also contribute to $w_{\rm int}$ evolving in time, so the sign of $\widetilde{K}_{\rm PS}$ does not necessarily mean the change in the local phase space internal energy density has the same sign. This is analogous to the fluid equation, given in Eq.~\ref{eq:IntEnerEq}, for which terms other than the pressure-strain interaction can change the local internal energy density, as has previously been pointed out \citep{Yang17, Du_ApJ_2018, Du_energy_2020_PRE, Cassak_PiD1_2022, Barbhuiya_PRE_2024}. In particular, although it is not the focus of the present study, as evident from Eq.~\ref{EQ:PSINTEQ2}, the collisional term $(1/2)mv^{\prime 2} C[f]$ can also change $w_{\rm int}$ in isolation and gives the volumetric collisional heating rate $\dot{Q}_{\rm coll}$ after integration over velocity space (as seen in Eq.~\ref{eq:IntEnerEq}), thereby capturing the irreversible increase in internal energy density. By contrast, a positive pressure-strain interaction is not by itself a measure of purely irreversible heating in weakly collisional plasmas.

\subsection{Decomposition of the alternative kinetic pressure-strain}
\label{sec:kpsdecomposition}

To investigate the phase-space origins of the decompositions of pressure-strain interaction that isolate the effects of normal vs.~shear flow effects, we develop decompositions of the alternative pressure-strain $\widetilde{K}_{\rm PS}$. The simplest decomposition of $\widetilde{K}_{\rm PS}$ from Eq.~\ref{eq:altKPS} is to separate the diagonal terms from the off-diagonal terms. For a 3-dimensional system, this decomposition is

\begin{equation}
\widetilde{K}_{\rm PS} = -\sum_{j=1}^{3} m v^{\prime 2}_j f \partial_j u_j - \sum_{j=1}^3 \sum_{k \neq j} m v^{\prime}_j v^{\prime}_k f \partial_j u_k, \label{eq:altKPS_decomp}
\end{equation}
where this expression is not using the Einstein summation convention.

The velocity space integral of the first term is $-\sum_{j=1}^3 P_{jj} \partial_j u_j$, which was defined as PDU in Ref.~\citep{Cassak_PiD1_2022}.  Thus, the first term is the phase-space analog of PDU, \textit{i.e.}, the part of the pressure-strain interaction that captures normal flow convergence/divergence physics.  
We therefore dub the first term ``kinetic PDU'', symbolically as $\widetilde{K}_{\rm PDU}$. In Cartesian coordinates, it is given by
\begin{equation}
    \widetilde{K}_{\rm PDU} = -\Bigl(m v_x^{\prime 2} \partial_x u_x 
+ m v_y^{\prime 2} \partial_y u_y 
+ m v_z^{\prime 2} \partial_z u_z \Bigr) f. \label{eq:KPDU}
\end{equation}

The velocity space integral of the second term in Eq.~\ref{eq:altKPS_decomp} is $- \sum_{j=1}^3 \sum_{k \neq j} P_{jk}\partial_j u_k$, which was defined as ${\rm Pi-D}_{\rm shear}$ in Ref.~\citep{Cassak_PiD1_2022}. Thus, the second term is the phase-space analog of ${\rm Pi-D}_{\rm shear}$, \textit{i.e.}, the part of the pressure-strain interaction that captures sheared flow physics \citep{Cassak_PiD1_2022}.  We analogously dub it ``kinetic ${\rm Pi-D}_{\rm shear}$,'' given symbolically as $\widetilde{K}_{\rm PiDS}$. In Cartesian coordinates, it is given by
\begin{equation}
\widetilde{K}_{\rm PiDS} = -\Bigl[m v_x^{\prime} v_y^{\prime} \left(\partial_x u_y+\partial_y u_x\right) + m v_y^{\prime} v_z^{\prime} \left(\partial_y u_z+\partial_z u_y\right) + m v_x^{\prime} v_z^{\prime} \left(\partial_x u_z+\partial_z u_x\right)\Bigr] f. \label{eq:KPiDS}
\end{equation}

Decomposing $\widetilde{K}_{\rm PS}$ into $\widetilde{K}_{\rm PDU}$ and $\widetilde{K}_{\rm PiDS}$ allows for an examination of the phase-space origins of PDU and ${\rm Pi-D}_{\rm shear}$, including the identification of particle populations that contribute most strongly to internal energy changes through normal (PDU) and sheared (${\rm Pi-D}_{\rm shear}$) flow effects. 
It was shown in Ref.~\citep{Conley_2024_PoP} that $\widetilde{K}_{\rm PS}$ can also be decomposed into the phase-space analogs of pressure dilatation $- \mathcal{P} (\nabla \cdot {\bf u})$ and Pi-D, which are called $\widetilde{K}_{\mathcal{P} \theta}$ and $\widetilde{K}_{\rm PiD}$, respectively.  For completeness, we provide these expressions in Cartesian coordinates in Appendix~\ref{appensec:B},  which are beneficial for determining the phase-space origins of compressible vs.~incompressible internal energy changes.

\subsection{Phase-space analog of the strain-rate tensor}

The phase-space analogs of the pressure-strain interaction and its decompositions reveal the phase-space origins of how different particle populations drive internal energy density evolution.  
While the phase-space analog of the pressure tensor enters these quantities, the strain-rate tensor $\partial_j u_k$ shows up only as a fluid variable. Thus, kinetically, only the bulk property of the strain-rate tensor matters for categorizing the phase-space origins of the internal energy changes, rather than any explicitly velocity-space resolved strain rate structure.

Although the strain-rate tensor is a bulk (fluid) quantity, it is ultimately determined by the kinetic (phase space) physics, which naturally raises the question: what are the phase-space origins of the strain-rate tensor itself? For example, in a plasma with multiple populations, one population may contribute disproportionately to the strain rate than another, but this information is lost in the fluid picture. Consequently, we develop the phase-space analog of the strain-rate tensor.

We introduce the ``kinetic strain-rate tensor" ${\bf K}_{\rm SR}$. The $jk$ element of ${\bf K}_{\rm SR}$ is 
\begin{equation}
    K_{{\rm SR}, {jk}} = \partial_j \left( \frac{f}{n} v_k \right) = \partial_j \left(\frac{f}{n} \right) v_k \label{eq:KSR}
\end{equation}
where $v_k$ passes through the spatial derivative because position and velocity are independent variables in kinetic theory. By construction, the velocity space integral of $K_{{\rm SR}, {jk}}$ is the corresponding strain-rate tensor element, {\it i.e.,} $\partial_j u_k = \int K_{{\rm SR}, {jk}} d^3v$.  The physical interpretation of ${\bf K}_{\rm SR}$ is that it is the strain-rate tensor per unit velocity-space volume, which is a function of phase-space coordinates.


\section{Numerical Simulations} 
\label{sec:simulation}

We utilize the massively parallel PIC code {\tt p3d} \citep{zeiler:2002} to perform a numerical simulation of symmetric antiparallel magnetic reconnection that is three-dimensional in velocity-space and two-dimensional in position-space to investigate the phase-space origins of internal energy density conversion through the kinetic pressure-strain. The simulation employs a relativistic Boris particle stepper for macro-particle evolution \citep{birdsall91a} and the trapezoidal leapfrog method for electromagnetic field evolution \citep{guzdar93a}. To enforce Poisson's equation, {\tt p3d} uses the multigrid method \citep{Trottenberg00}. Periodic boundary conditions are applied in both spatial directions.

All the simulation data in the present study are in \texttt{p3d} normalized units. 
Magnetic fields are normalized to the initial asymptotic magnetic field strength $B_0$.
Velocity scales are normalized to the Alfv\'en speed $c_{A0} = B_0/(4 \pi m_i n_0)^{1/2}$ based on the reference number density $n_0$, and where $m_i$ the ion mass.
Time scales are normalized to the inverse ion cyclotron frequency $\Omega_{ci0}^{-1}= (q_i B_{0} / m_{i} c)^{-1}$, where $q_i$ is the ion charge, and $c$ is the speed of light.
Length scales are normalized to the ion inertial scale $d_{i0} = c/\omega_{pi0}$, where $\omega_{pi0} = (4 \pi n_0 q_i^2 /m_i)^{1/2}$ is the ion plasma frequency.
Temperatures are normalized to $m_i c_{A0}^2/k_B$. 

The simulation in this study was previously used in another study \citep{Barbhuiya_PiD3_2022}. Unrealistic values for the speed of light ($c=15$) and the electron-to-ion mass ratio ($m_e/m_i = 0.04$) are employed for computational efficiency. While the choice of mass ratio is sufficient for the present study since a more realistic mass ratio alters the amplitudes of the quantities studied (while preserving the qualitative features \citep{Edyvean2024ApJ}), we do note that realistic scale separation can influence electron-scale structure and wave/instability activity in thin current sheets. The grid scale is $\Delta=0.0125$, which is smaller than the system's smallest length scale, the electron Debye length $\lambda_{De}=0.0176$.
Similarly, we use a particle time step  $\Delta t=0.001$, and the electromagnetic field time step is half of the particle time step.
The electric field is cleaned every 10 particle time steps.  
The simulation domain size is $L_x \times L_y =  12.8 \times 6.4$, which equals $1024 \times 512$ grid cells, that are initialized with 25,600 weighted particles per grid cell (PPG).
The domain is small enough that the system reaches the growth phase with ions partially recoupling to the reconnected magnetic field, but not large enough to reach a steady-state phase \citep{Barbhuiya_phases_2025}. 
We analyze the simulation data at $t = 13$, when the time rate of change of reconnection rate is the highest, dubbed the ``peak growth-time" \citep{Barbhuiya_phases_2025}. We analyze the pressure-strain interaction only of the electrons and focus on the lower current sheet.

\begin{figure*}[t]
    \centering
    \includegraphics[width=\textwidth]{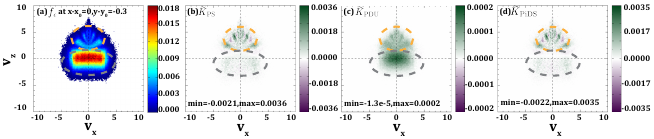}
    \caption{
    (a) Electron VDF $f_e$, (b) $\widetilde{K}_{\rm PS}$, (c) kinetic PDU $\widetilde{K}_{\rm PDU}$, and (d) kinetic ${\rm Pi-D}_{{\rm shear}}$ $\widetilde{K}_{\rm PiDS}$ near the upstream edge of the EDR at $(x-x_0,y-y_0)=(0,-0.3)$.  Each is reduced to the $v_x-v_z$ plane. The gray and gold ovals highlight the upstream electrons drifting towards the EDR and the demagnetized electrons undergoing Speiser motion, respectively.} 
    \label{fig:upstreamEDRKPSdecomp}
\end{figure*}

At initialization, the magnetic field profile is defined using a double ${\rm tanh}$ function, $B_x(y) = \tanh{\left[(y-L_y/4)/w_0\right]} - \tanh{\left[(y-3L_y/4)/w_0\right]} - 1$, with no (out-of-plane) guide field, so $B_z = 0$. Here, $w_0 = 0.5$ is the half-thickness of the current sheet. The electron and ion density profiles are given by $n(y) = \{1/[2(T_{e}+T_{i})] \} \{ {\rm sech}^{2}\left[(y-L_y/4)/w_0\right] + {\rm sech}^{2}\left[(y-3L_y/4)/w_0\right] \} + n_{up}$, where the asymptotic upstream plasma density is $n_{up} = 0.2$, and the difference between the peak current sheet number density and $n_{up}$ defines the reference density $n_0$. The electron and ion temperatures are uniform and set to $T_e = 1/12$ and $T_i = 5T_e$, respectively, with their out-of-plane drift velocities $u_{\sigma,z}$ constrained by $u_{e,z}/T_e = -u_{i,z}/T_i$. The asymptotic total plasma beta (the ratio of the total plasma gas pressure to the magnetic pressure) is 0.2. Both species are initialized as single drifting Maxwellian velocity distribution functions rather than the two-population model used for the Harris sheet kinetic equilibrium \citep{Harris62}. Reconnection is triggered by perturbing the magnetic field with a magnetic field perturbation given by $\delta B_x = -B_{pert} \sin \left(2 \pi x/L_x \right) \sin\left(4\pi y/L_y \right)$ and $\delta B_y =  B_{pert} \left[L_y/(2 L_x)\right] \cos\left(2 \pi x/L_x\right) \left[1-\cos\left(4\pi y/L_y\right)\right]$, with $B_{pert} = 0.05$. 

For plots of electron velocity distribution functions (VDFs) presented in this study, we employ a domain of size $5\Delta \times 5\Delta =  0.0625 \times 0.0625$ centered at the location of interest to collect particles that are then binned with a velocity-space bin of size 0.1 in all velocity directions to produce the full 3-dimensional velocity distribution function. These 3-D velocity distribution functions are then reduced by integrating out one velocity-space direction $v_j$ and plotted in $v_k-v_l$ planes (where $j,~k,~l=x,~y,~z$). Such reduced velocity distribution functions, with one velocity dimension integrated out, are normalized to $n_0/c^2_{A0}$.

Because the VDFs employ a spatial domain of 5 cells, we smooth all electron fluid data, \textit{e.g.,} number density, once over a width of five spatial grid cells in each position-space direction using the native IDL function SMOOTH, which employs a boxcar average. 
For fluid quantities that require spatial derivatives, \textit{e.g.,} strain-rate tensor elements, we first smooth bulk flow velocities over a width of five spatial cells, then carry out the spatial derivatives, and then smooth the results again over five cells.

\section{Results} 
\label{sec:results}

We now analyze the electron $\widetilde{K}_{\rm PS}$ and its decomposition into $\widetilde{K}_{\rm PDU}$ and $\widetilde{K}_{\rm PiDS}$ and the kinetic strain-rate tensor ${\bf K}_{\rm SR}$ 
near the electron diffusion region (EDR) to determine the phase space structures that lead to the pressure-strain interaction and the strain-rate tensor, respectively. We obtain information about the contributions from different populations in these multi-population VDFs and contrast the dominant populations.

We first briefly review the bulk properties of the electron pressure-strain interaction, which was studied in depth in Ref.~\citep{Barbhuiya_PiD3_2022}. In this reference, Fig.~1 contains all six independent pressure tensor elements and the six non-zero elements of the electron strain-rate tensor, and Fig.~2 contains plots of the pressure-strain interaction and its decompositions.
At the time of interest $t=13$, the electron diffusion region (EDR) extends from approximately  $|x-x_0| \lesssim 2$, $|y-y_0| \lesssim 0.35$.   

Close to the upstream edge of the EDR $(x-x_0,y-y_0)=(0,-0.3)$, PDU is weakly positive due to the slowing down of the electron inflow manifesting as flow convergence, while the out-of-plane ($z$) electron flow shears in the inflow ($y$) direction, and $P_{yz}$ is the dominant off-diagonal pressure tensor element, resulting in strong positive ${\rm Pi-D}_{\rm shear}$. 
Near the X-line $(x-x_0,y-y_0)=(0,0)$, PDU is moderately negative due to the diverging electron outflows resulting in expansion, while ${\rm Pi-D}_{\rm shear}$ is negligible because the bulk flow shear is weak due to symmetry.
Downstream of the X-line near the outflow edge of the EDR $(x-x_0,y-y_0)=(-1.4,0)$, electrons run into a pre-existing current sheet and the growing island, which leads to a converging bulk flow producing a strong positive PDU.
Here, out-of-plane ($z$) electron flow shears in the outflow ($x$) direction resulting in weak positive ${\rm Pi-D}_{\rm shear}$. We thus find that shear effects are most important near the upstream edge of the EDR, while normal flow effects are most important near to and downstream of the X-line \citep{Barbhuiya_PiD3_2022}.

\begin{figure*}
    \centering
    \includegraphics[width=\textwidth]{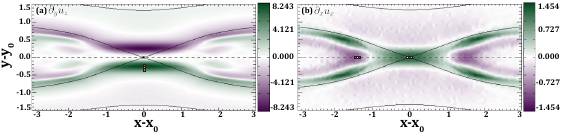}
    \caption{Two strain-rate tensor elements centered at the X-line $(x_0,y_0)$. (a) $\partial_y u_z$, and (b) $\partial_x u_x$. Colored boxes denote the spatial regions from which particles are collected and binned to produce 3-D VDFs as detailed in Sec.~\ref{sec:simulation}. 
    Representative in-plane magnetic field lines are shown in black.}
    \label{fig:SRelements}
\end{figure*}
 
The three locations discussed in the previous paragraph — upstream edge of the EDR, X-line, and outflow edge of the EDR — represent three distinct regions of the fluid dynamics of reconnection. However, they also represent distinct environments kinetically. 
In what follows, we perform an in-depth analysis of the phase-space origins of the pressure-strain interaction and kinetic strain-rate tensor at these three locations. 
For reference, we present the reduced electron VDF in each of the Cartesian directions at three locations considered here
in Appendix \ref{appensec:C}. 
For the analysis that follows, we highlight the reduced VDFs in the $v_x-v_z$ plane because they are qualitatively easier to visualize the multiple electron populations in the VDFs.

\subsection{Near the upstream edge of the EDR} \label{subsec:result_upstreamEDR}

We start by presenting the reduced electron VDF at $(x-x_0,~y-y_0)=(0,-0.3)$
in Fig.~\ref{fig:upstreamEDRKPSdecomp}(a). 
We first identify different populations present in the electron VDF:

\begin{enumerate}
    \item One population, denoted by the gray dashed oval, comes from the upstream region and is drifting towards the EDR. The phase-space density from this population is elongated in the $v_\parallel \approx v_x$ direction, where $\parallel$ is the direction along the local magnetic field. This population is itself comprised of two populations, trapped electrons with low $|v_x|$, $|v_y|$, and $|v_z|$, and passing electrons at high $|v_x|$, low $|v_y|$, and low $|v_z|$ \citep{Egedal13}. 

    \item The second population, highlighted by the gold dashed oval in Fig.~\ref{fig:upstreamEDRKPSdecomp}(a), consists of electrons that become demagnetized after encountering the magnetic field reversal    \citep{Ng_2011_PRL} and perform Speiser motion \citep{Speiser65}. These contribute to the various striations based on the number of bounces     \citep{Ng_2011_PRL, Ng_2012_PoP} while being accelerated in the $z$ direction by the reconnection electric field \citep{Bessho_2018_GRL}. 
\end{enumerate} 

We now analyze the phase-space origin of the pressure-strain interaction and its decompositions at the specified location.  Figure~\ref{fig:upstreamEDRKPSdecomp}(b) shows $\widetilde{K}_{\rm PS}$ reduced into the $v_x-v_z$ plane. The gray and gold ovals are included to guide the eye.
Direct inspection of $\widetilde{K}_{\rm PS}$ reveals that it is relatively small in magnitude for the drifting population (gray oval), so the dominant contribution is from the demagnetized population (gold oval). The demagnetized population has structures with alternating regions in velocity space of positive and negative $\widetilde{K}_{\rm PS}$ due to the Speiser motion, but the net contribution to the pressure-strain interaction is positive.
Interestingly, the phase-space density of the demagnetized population is smaller than that of the drifting population by about a factor of three, yet the former population has the dominant contribution to $\widetilde{K}_{\rm PS}$ by a couple of orders of magnitude. 
The velocity-space regions with positive and negative $\widetilde{K}_{\rm PS}$ contribute to local increases and decreases in $w_{\rm int}$, respectively, in isolation, with the net effect of $\widetilde{K}_{\rm PS}$ being positive, which is why $\mathcal{E}_{\rm int}$ increases due to positive pressure-strain interaction.

We now assess the terms in the decomposition of kinetic pressure-strain to ascertain whether it arises as a result of normal or sheared flow. Figure~\ref{fig:upstreamEDRKPSdecomp}(c) and (d) give $\widetilde{K}_{\rm PDU}$ and $\widetilde{K}_{\rm PiDS}$, respectively, reduced into the $v_x-v_z$ plane. We note that the color bars on the two panels are different. 
Indeed, the structure of $\widetilde{K}_{\rm PiDS}$ is nearly identical to $\widetilde{K}_{\rm PS}$, so the pressure-strain dominantly comes from sheared flow effects. This finding is consistent with the fluid picture result as ${\rm Pi-D}_\text{shear} = \int  \widetilde{K}_{\rm PiDS} d^3v \approx 0.012$ is larger than ${\rm PDU}  = \int \widetilde{K}_{\rm PDU} d^3v\approx 0.003$ at this location.
The drifting electron population has a relatively stronger $\widetilde{K}_{\rm PDU}$ than the demagnetized population and almost no $\widetilde{K}_{\rm PiDS}$. This result is because the trapped and passing particles at this location lead to elongated distributions in $v_x$ leading to a non-zero ${\rm Pi-D}_\text{normal}$, but the population has very little flow shear.
 
We turn to analyze the phase-space origin of the strain-rate tensor at this location. 
There are six elements that are non-zero in 2-D as $\partial_z \rightarrow 0$ (not shown).  
We find that the dominant kinetic strain-rate tensor element is $K_{{\rm SR},{yz}}=\partial_y(f/n)v_z$. 
The strain-rate tensor element $\partial_y u_z$ is plotted in Fig.~\ref{fig:SRelements}(a), centered at the X-line location $(x_0,y_0)$ for the lower current sheet and containing only a portion of the full computational domain, with representative in-plane magnetic field lines shown in black.  

\begin{figure}
    \includegraphics[width=2.5in]{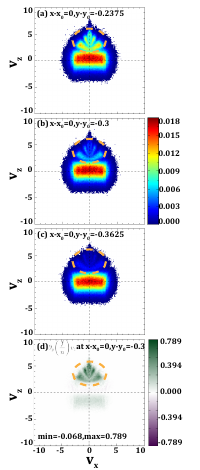}
    \caption{Electron VDF near the upstream edge of the EDR, reduced in the $v_x-v_z$ plane at (a) $(x-x_0,y-y_0)=(0,-0.2375)$, (b) $(x-x_0,y-y_0)=(0,-0.3)$, and (c) $(x-x_0,y-y_0)=(0,-0.3625)$. (d) Reduced kinetic strain-rate tensor element $K_{{\rm SR},yz}$ in the $v_x-v_z$ plane.}    
    \label{fig:upstreamEDR}
\end{figure}

Figure~\ref{fig:upstreamEDR}(a)-(c) shows the spatial variation of the reduced VDF in the $v_x-v_z$ plane required for calculating $\partial_y (f/n)$. The spatial regions used to calculate these VDFs are shown in  Fig.~\ref{fig:SRelements}(a), where $(x-x_0,y-y_0)=(0,-0.2375)$ is the center of the red box, $(0,-0.3)$ is the center of the orange box (the location of interest), and $(0,-0.3625)$ is the center of the green box, shown in Fig.~\ref{fig:upstreamEDR}(a)-(c), respectively. 
The reduced VDFs at the two neighboring cells in the $y$ direction are included to highlight how the VDF changes in $y$.

The velocity space structure of $K_{{\rm SR},yz}$ resulting from finite differences of the VDFs, reduced into the $v_x-v_z$ plane, is plotted in Fig.~\ref{fig:upstreamEDR}(d). The dominant contribution is from the demagnetized population, marked by the gold oval. Physically, as highlighted by the gold ovals in Fig.~\ref{fig:upstreamEDR}(a)-(c), the number of particles undergoing Speiser motion increases closer to the X-line because more electrons sample the reversed magnetic field during their gyromotion. Thus, $f/n$ increases as one approaches the X-line. 
The bulk flow $u_z$ also comes mostly from the demagnetized electrons, which for this current sheet is positive. This result explains the sign and structure of the strain-rate tensor element $\partial_y u_z$ in Fig.~\ref{fig:SRelements}(a). 
The contribution of the drifting population is a few orders of magnitude smaller than that of the demagnetized population.

In summary, we find that at the location $(x-x_0,y-y_0)=(0,-0.3)$ near the upstream edge of the EDR, the phase-space origin of the pressure-strain interaction is predominantly from the demagnetized electrons, with the drifting population having only a small effect. The dominant physical cause is sheared flow, which is also predominantly caused by the demagnetized electrons. This dominance is the case even though the phase-space density of this population is lower than that of the drifting population at this location. 

\subsection{At the X-line}\label{subsec:res_xline}

\begin{figure*}[t]
    \centering
    \includegraphics[width=\textwidth]{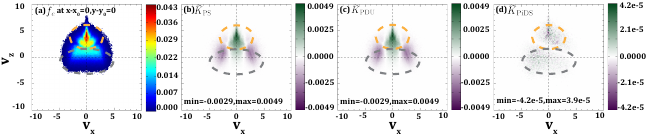}
    \caption{Same as Fig.~\ref{fig:upstreamEDRKPSdecomp}, but at the X-line $(x-x_0,y-y_0) = (0,0)$. The gray ovals highlight populations of electrons drifting towards and away from the X-line, and the gold ovals highlight populations of the demagnetized electrons undergoing Speiser motion.}
    \label{fig:XlineKPSdecomp}
\end{figure*}

\begin{figure*}[t]
   \centering
    \includegraphics[width=\textwidth]{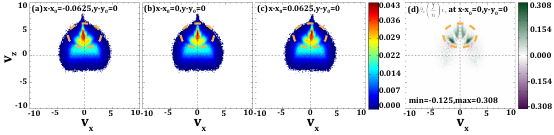}
   \caption{Analogous to Fig.~\ref{fig:upstreamEDR}, except the electron VDF is near the X-line at an $(x-x_0,y-y_0)$ of (a) $(-0.0625,0)$, (b) $(0,0)$, and (c) $(0.0625,0)$. (d) Reduced kinetic strain-rate tensor element $K_{{\rm SR},xx}$ in the $v_x-v_z$ plane at $(x-x_0,y-y_0)=(0,0)$.}
   \label{fig:Xline}
\end{figure*}

We repeat the analysis in the previous subsection at the X-line. 
Fig.~\ref{fig:XlineKPSdecomp}(a) shows the reduced electron VDF at  $(x-x_0,~y-y_0)=(0,~0)$. We designate the populations in the electron VDF as follows. 

\begin{enumerate}
    \item We group the electrons drifting inward towards the X-line from above $(y-y_0>0)$ and below $(y-y_0<0)$, and those drifting outwards away from the X-line to the left $(x-x_0<0)$ and right $(x-x_0>0)$, marked by the gray dashed oval.
    
    \item Two demagnetized electron populations from either upstream side of the X-line with discrete levels whose constituents are undergoing Speiser motion \citep{Bessho_GRL_2014}, marked by the gold dashed oval. 
\end{enumerate}

The reduced $\widetilde{K}_{\rm PS}$ is shown in Fig.~\ref{fig:XlineKPSdecomp}(b). There are two regions of phase space with a negative $\widetilde{K}_{\rm PS}$ associated with the drifting population and one region of strong positive $\widetilde{K}_{\rm PS}$ associated with the demagnetized electrons. The negative $\widetilde{K}_{\rm PS}$ is associated with electrons which are redirected in the $\pm x$ directions into the outflow jet. Since the bulk flow speed $u_x$ increases with increasing downstream distance away from the X-line, this population is expanding. 
The positive $\widetilde{K}_{\rm PS}$ is due to electrons being accelerated in the out-of-plane direction by the reconnection electric field as they undergo Speiser motion \citep{Speiser65}, which spreads the distribution in $v_z$. 
The combined effect of all populations, given by the velocity space integral of $\widetilde{K}_{\rm PS}$, is negative, meaning that the pressure-strain interaction is negative at the X-line, as seen in Fig.~2 in Ref.~\citep{Barbhuiya_PiD3_2022}. Thus, at the X-line at the time shown here, the expansion of the electrons into the outflow regions has a greater net negative effect than the demagnetized electrons have a positive effect. This result is an interesting example of how the phase space picture reveals that even at a single fluid element, there may be multiple populations having opposite contributions, which could not be ascertained from the fluid picture.

The quantities $\widetilde{K}_{\rm PDU}$ and $\widetilde{K}_{\rm PiDS}$ making up the decomposition of kinetic pressure-strain are plotted in Fig.~\ref{fig:XlineKPSdecomp}(c) and (d), respectively. We immediately see that the sheared flow portion given by $\widetilde{K}_{\rm PiDS}$ is nearly zero, two orders of magnitude smaller than $\widetilde{K}_{\rm PDU}$. Physically, the flow into and out of the X-line region is purely normal due to symmetry, so the bulk flow shear is nearly zero. 
At this location, ${\rm PDU} = \int \widetilde{K}_{\rm PDU} d^3v \approx -0.008$, which dominates over ${\rm Pi-D}_\text{shear} = \int \widetilde{K}_{\rm PiDS} d^3v \approx 2 \times 10^{-5}$ \citep{Barbhuiya_PiD3_2022}. 
The outflowing populations have a lower phase-space density by a factor of six than the demagnetized population, but they produce the dominant contribution to the electron pressure-strain interaction.

\begin{figure*}[t]
    \centering
    \includegraphics[width=\textwidth]{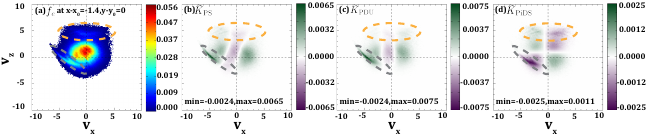}
    \caption{Same as Fig.~\ref{fig:upstreamEDRKPSdecomp}, but near the downstream edge of the EDR $(x-x_0,y-y_0) = (-1.4,0)$. The gray and gold ovals highlight electrons that are remagnetizing and an already remagnetized population, respectively.}
    \label{fig:outflowEDRKPSdecomp}
\end{figure*}

To ascertain the phase-space origin of the strain-rate tensor at the X-line, we first find that the dominant element at the X-line is $\partial_x u_x$, which is plotted in Fig.~\ref{fig:SRelements}(b).
The associated element of the kinetic strain-rate tensor is
$K_{{\rm SR},{xx}}=\partial_x(f/n)v_x$. 
The VDFs that go into the calculation of this element are plotted in Fig.~\ref{fig:Xline}(a)-(c), where the VDF at $(x-x_0,y-y_0)=(-0.0625,0)$ is from the white box in Fig.~\ref{fig:SRelements}(b), $(0,0)$ is the red box (the location of interest), and $(0.0625,0)$ is from the light blue box, respectively. The resultant $K_{{\rm SR},{xx}}=\partial_x(f/n)v_x$ is in Fig.~\ref{fig:Xline}(d). 

The phase space structure of $K_{{\rm SR},xx}$ reveals that its dominant contribution is from the demagnetized population (marked by the gold oval). 
Physically, this result comes about because on either side of the X-line, the demagnetized populations are slightly rotated by the reconnected magnetic field $B_y$ \citep{Shuster14, shuster_spatiotemporal_2015, BarbhuiyaRing22, Ng_2012_PoP,Bessho_GRL_2014}. This rotation is visible in the small leftward and rightward rotation seen in the demagnetized population (gold ovals) in Figs.~\ref{fig:Xline}(a) and (c) relative to (b), respectively. The rotation induces a spatial variation in $x$ in the phase-space density at the X-line. In the fluid picture, this rotation decreases $u_z$ while producing a non-zero $u_x$ that increases when moving away from the X-line.  
From Fig.~\ref{fig:Xline}(d), we find that $K_{{\rm SR},xx}$ is predominantly positive, explaining, kinetically, why $\partial_x u_x>0$ at this location. 

In summary, we find that at the X-line, the population providing the predominant contribution to the negative pressure-strain interaction is the electrons that are being directed into the outflow. Interestingly, the demagnetized population at the same location contributes positively to the pressure-strain interaction even though it does not dominate. 
Intriguingly, the population that gives the dominant contribution to the strain-rate tensor is the demagnetized population, which is different than the population giving the dominant pressure-strain interaction. 
This finding underscores the importance of studying the phase-space origins of the strain-rate tensor since the population giving the dominant pressure-strain interaction need not be the same as the population giving the dominant contribution to the strain-rate tensor.

\subsection{Near the outflow edge of the EDR}

\begin{figure*}[t]
   \centering
    \includegraphics[width=\textwidth]{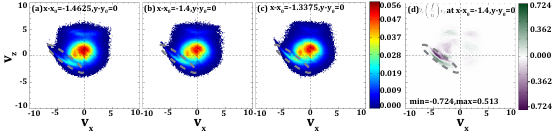}
   \caption{Analogous to Fig.~\ref{fig:upstreamEDR}, except near the downstream edge of the EDR with VDFs at an $(x-x_0,y-y_0)$ of (a) $(-1.4625,0)$, (b) $(-1.4,0)$, and (c) $(-1.3325,0)$. (d) Reduced kinetic strain-rate tensor element $K_{{\rm SR},xx}$ in the $v_x-v_z$ plane at $(x-x_0,y-y_0)=(-1.4,0)$.}
   \label{fig:downstreamEDR}
\end{figure*}

Finally, we consider the outflow edge of the EDR where relatively large values of PDU in Ref.~\citep{Barbhuiya_PiD3_2022} are located. 
The reduced VDF at $(x-x_0,~y-y_0)=(-1.4,0)$ is shown in Fig.~\ref{fig:outflowEDRKPSdecomp}(a) and we identify the following populations.

\begin{enumerate}
    \item A core electron population.  This population has some substructure (not shown), but the substructure is not important for the purposes of this study, so we ignore it.
    
    \item An electron population that has remagnetized in the reconnected magnetic field $B_y$. \citep{shuster_spatiotemporal_2015,Bessho_GRL_2014}. 
    This population appears as an arc at the highest $v_z>0$ values and is marked by a gold dashed oval. (See, \textit{e.g.,} at the arc with $v_y<0$ and small $|v_x|$ in Fig.3 4E of Ref.~\citep{shuster_spatiotemporal_2015}.)

    \item An electron population that is in the process of magnetizing, marked by the gray oval. These electrons once formed the striations near the X-line as they performed Speiser orbits. 
    At this downstream location, these electrons are being rotated by $B_y$ and appear as incomplete crescents \citep{Bessho_GRL_2014,shuster_spatiotemporal_2015}. 
    (See, \textit{e.g.,} at the arcs with $v_x >0$ and $v_y>0$ in Fig.3 4D and 4E of Ref.~\citep{shuster_spatiotemporal_2015}.)
\end{enumerate}

To identify the electron populations that contribute dominantly to the electron pressure-strain interaction near the outflow edge of the EDR, we analyze the reduced $\widetilde{K}_{\rm PS}$ shown in Fig.~\ref{fig:outflowEDRKPSdecomp}(b).  
All three populations contribute to the kinetic pressure-strain. The amplitude is strongest for the incomplete crescent population, 
and the net pressure-strain interaction is positive at this location.
Physically, the cause of the positive pressure-strain interaction is that electrons undergoing Speiser orbits near the X-line [the population in the gold oval in Fig.~\ref{fig:XlineKPSdecomp}(a)] get rotated by the reconnected magnetic field into the outflow direction.  Electrons with $v_z$ slower than half the Alfv\'en speed complete gyro-orbits inside the EDR, and become part of the magnetized populations [the population in the gray oval in Fig.~\ref{fig:outflowEDRKPSdecomp}(a)].  Electrons with $v_z$ faster than half the Alfv\'en speed make up the incomplete crescent population at the downstream edge of the EDR.
At the downstream edge of the EDR, the electrons in the incomplete crescent population are still being rotated by the reconnected magnetic field such that their $|v_z|$ and $|v_x|$ are lowered with increasing distance from the X-line \citep{Bessho_GRL_2014,shuster_spatiotemporal_2015}.  Thus, their contribution to $\partial u_x / \partial x$ is negative (see Fig.~1(m) in Ref.~\citep{Barbhuiya_PiD3_2022}). Since $P_{xx}$ is positive (see Appendix \ref{appensec:C}), their $\widetilde{K}_{\rm PS}$ is positive.

The phase-space origins of the decomposed kinetic pressure-strain are revealed in Fig.~\ref{fig:outflowEDRKPSdecomp}(c)-(d), showing the reduced $\widetilde{K}_{\rm PDU}$ and $\widetilde{K}_{\rm PiDS}$, respectively. The incomplete crescent population (gray ovals) produces the largest signal in both quantities, although interestingly, they are of opposite signs. Because the maximum value of $\widetilde{K}_{\rm PDU}$ exceeds the maximum value of $\widetilde{K}_{\rm PiDS}$ by a factor of three, the net contribution to the kinetic pressure-strain is positive. 
Physically, as discussed in the previous paragraph, the dominant term is due to the normal flow effect $-m v_{x}^{\prime 2} f\partial u_x / \partial x$. $\widetilde{K}_{\rm PiDS}$ is negative at this location because $v_z$ is increasing in $x$, as discussed in the previous paragraph, and $P_{xz}$ is positive [see Fig.~1(e) in Ref.~\citep{Barbhuiya_PiD3_2022}], so $\widetilde{K}_{\rm PiDS} <0$. The other populations have much smaller contributions to kinetic pressure-strain.  
At this location, ${\rm PDU} = \int \widetilde{K}_{\rm PDU} d^3v \approx 0.05$, almost an order of magnitude larger than ${\rm Pi-D}_\text{shear} = \int \widetilde{K}_{\rm PiDS}d^3v \approx 0.007$, consistent with the fluid picture \citep{Barbhuiya_PiD3_2022}.

To analyze the phase-space origin of the strain-rate tensor at $(x-x_0,y-y_0)=(-1.4,0)$, we first find that the dominant element of the strain-rate tensor is $\partial u_x / \partial x$, which was the dominant term at the X-line and is plotted in Fig.~\ref{fig:SRelements}(b).
Fig.~\ref{fig:downstreamEDR}(a)-(c) is analogous to Fig.~\ref{fig:Xline} but for $(x-x_0,y-y_0)=(-1.4,0)$.  
The VDFs are from $(x-x_0,y-y_0)=(-1.4625,0)$, the magenta box in Fig.~\ref{fig:SRelements}(b), $(-1.4,0)$ is the yellow box (the location of interest), and $(-1.3375,0)$ is the blue box, respectively. 

Figure~\ref{fig:downstreamEDR}(d) gives the finite difference result for the reduced kinetic strain-rate tensor element $K_{{\rm SR},xx} = \partial_x (f/n) v_x$. 
Inspection of $K_{{\rm SR},xx}$ reveals that the incomplete crescent population (gray oval) is the dominant contribution to $K_{{\rm SR},xx}$. 
Physically, electrons remagnetize as they interact with the reconnected magnetic field, whose strength increases with distance from the X-line. 
The remagnetization process decreases $u_x$ with distance from the X-line \citep{Shuster14,shuster_spatiotemporal_2015,Bessho_GRL_2014,BarbhuiyaRing22}, manifesting as normal flow convergence $\partial_x u_x < 0$.
We note that $K_{{\rm SR},xx}$ has a bipolar signature for the dominant population, but because the negative signal is stronger, the strain-rate tensor element $\partial_x u_x = \int K_{{\rm SR},xx} d^3v$ is negative, as found in the fluid picture \citep{Barbhuiya_PiD3_2022}. More significantly, the incomplete crescent population has a factor of two smaller phase-space density, but contributes most dominantly to the strain-rate tensor.

In summary, at this location near the outflow edge of the EDR, we find that the incomplete crescent population dominates both the strain-rate tensor and the pressure-strain interaction, primarily via normal flow convergence effects. This dominance is the case even though this population is in the minority of the total distribution of particles there. This result is counterintuitive. It was previously supposed that the contribution to the positive pressure-strain interaction at this location just downstream of the EDR came from the bulk outflow jet impinging on the island \citep{Barbhuiya_PiD3_2022}, but the phase-space analysis reveals this is not the case. 

\section{Discussion and Conclusion} 
\label{sec:conclusion}

In a previous study \citep{Conley_2024_PoP}, the phase-space analog of the pressure-strain interaction, dubbed the kinetic pressure-strain $K_{\rm PS}$ (or an alternative form $\widetilde{K}_{\rm PS}$), was derived and used to study the phase-space origin of energy conversion into internal energy resulting from bulk compression, deformation, and/or flow shear. The focus was on using kinetic pressure-strain to study wave-particle interactions as a means to distinguish between the effects of the resonant and non-resonant particles driving the energization during Landau damping.  In this study, we delve further into the nature of the alternative kinetic pressure-strain and elaborate on many other processes for which the quantity is useful.

First, we revisit the derivation of $\widetilde{K}_{\rm PS} = -m v_j^\prime v_k^\prime f \partial_j u_k$ in the laboratory rest frame, and derive it from the kinetic internal energy density equation.  This approach has the advantage of also identifying other terms that describe the evolution of the phase space internal energy density $w_{{\rm int}}$.  We show that the decomposition into the effects due to normal flow (the phase-space analog of PDU) and flow shear (the phase-space analog of ${\rm Pi-D}_{{\rm shear}}$) follows naturally. We then argue that a more complete picture of the phase-space origins of the pressure-strain interaction benefits from documenting the phase-space origins of the strain-rate tensor itself. Thus, we introduce 
the phase-space analog of the strain-rate tensor, which we call ${\bf K}_{\rm SR}$.
We reason that the phase-space approach, in addition to being useful for distinguishing between resonant and non-resonant particle populations in a resonant interaction, is also useful in multi-population plasmas that often arise in weakly collisional plasmas to disambiguate which population or populations are contributing the most to the fluid behavior.

We use the well-studied process of collisionless anti-parallel magnetic reconnection, for which multi-population plasmas are known to exist, to exemplify the utility of this approach. We employ an in-depth analysis of the velocity distribution functions produced in kinetic particle-in-cell simulations near the upstream edge of the electron diffusion region, near the X-line, and near the outflow edge of the electron diffusion region.  We find several interesting results coming from the complicated phase space behavior: (1) the alternative pressure-strain interaction successfully disambiguates the roles of different populations, and its decomposition into $\widetilde{K}_{{\rm PDU}}$ and $\widetilde{K}_{{\rm PiDS}}$ reveals whether the effect is due to normal or sheared bulk flow; (2) the kinetic pressure-strain can clarify when there are multiple populations and the contribution from two (or more) populations opposes each other at the same physical location;
(3) the dominant population causing the pressure-strain interaction can be a population with a relatively small phase-space density; (4) while the population that has the dominant contribution to the pressure-strain interaction may be the same population that gives the dominant contribution to the strain-rate tensor, there are also circumstances for which different populations can be dominant for the two different terms.
The kinetic strain-rate tensor identifies which population produces local flow dilatation and deformation, while the kinetic pressure-strain identifies which population actually mediates the local change of internal energy density through that dilatation/deformation. Thus, in a multi-component VDF, the particles that set the local flow kinematics need not be the particles that dominate the local energetics. 
This distinction is realized most clearly near the reconnection site, where we find that electrons being redirected into the outflow cause the net negative pressure-strain interaction, whereas the demagnetized Speiser populations produce the dominant strain-rate tensor.
When used together, KPS and KSR therefore indicate whether the same population shapes the change in bulk flows and leads to local energy density evolution, or if these roles are performed by different populations. 

For the magnetic reconnection process studied here, we reproduce some expected but also some potentially surprising results. 
At the physical location we considered near the upstream edge of the EDR, the electrons undergoing Speiser orbits and getting accelerated by the reconnection electric field give the biggest contribution to the pressure-strain interaction, which is positive. The decomposition of KPS into kinetic ${\rm PDU}$ and kinetic ${\rm Pi-D}_\text{shear}$ shows that this contribution is primarily associated with sheared flow effects.
The same Speiser electron population dominates the bulk velocity shear because the number of electrons undergoing Speiser orbits increases as one gets closer to the X-line. 
In contrast, near the X-line itself, the negative pressure-strain interaction is predominantly provided by the electrons that are redirected into outflow jets, while the electrons undergoing Speiser orbits predominantly provide the strain rate. Thus, near the reconnection site, the kinetic decomposition shows that the pressure-strain interaction is instead dominated by normal flow effects/expansion rather than shear.
At the position-space location we considered near the downstream edge of the EDR, an incomplete crescent population of remagnetizing electrons provides the dominant contribution to the net positive pressure-strain interaction, primarily through normal flow convergence, and the same population is responsible for the dominant strain rate. 
In all three of the selected locations, the populations listed here dominate despite being a relatively small phase-space density. 
Taken together, the phase space level analyses at the upstream edge of the EDR, X-line, and outflow edge of the EDR demonstrate that the relationship between pressure–strain interaction and strain rate is not uniform across the EDR.
They also show that the physical character of the pressure-strain interaction changes across the EDR: it is shear-dominated near the upstream edge, but normal-flow dominated near and downstream of the reconnection site. Thus, the decomposition of KPS extends the fluid-level interpretation of Ref.~\citep{Barbhuiya_PiD3_2022} by identifying which velocity-space populations mediate the corresponding sheared-flow or normal-flow contribution, thereby highlighting
the importance of a phase-space approach for understanding energy evolution in magnetic reconnection.

A broader implication of these results is that the local phase-space density alone is not a reliable indicator of which particle population is most important for energy evolution in a multi-population velocity distribution function. The population that dominates KPS (and often KSR) can have relatively small phase-space density, but its relative contribution to the phase space energy evolution gets amplified since KPS is weighted not only by $f$, but also by the peculiar velocity factors $v'_j$ and the local (fluid) strain-rate tensor element $\partial_j u_k$. Similarly, KSR is instead sensitive to spatial variations in $f/n$. Thus, a population that is visually subdominant in $f$ can dominate the fluid pressure-strain interaction or the fluid strain-rate tensor if it occupies regions of sufficiently large $|{\bf v}'|$ compared to the dominant population(s) in $f$, has strong spatial gradients, and/or contributes disproportionately to the pressure tensor. In the present study, this occurs for the demagnetized Speiser populations near the upstream edge of the EDR, the outflowing electron population near the X-line, and the incomplete crescent population near the outflow edge of the EDR. These findings underscore the need for phase-space-resolved analyses of energy evolution in weakly collisional multi-population plasmas.

The in-depth use of the terms in this study to understand the phase-space origins of energy evolution proves useful to study magnetic reconnection, but it should be useful for other fundamental physical processes.  It has already been shown that a phase-space analysis could be useful for wave-particle interactions \citep{Conley_2024_PoP}. We expect it will be useful for magnetized turbulence in weakly collisional or collisionless systems, and especially for collisionless shocks.  The key quantities in question are relatively straightforward to obtain using other particle-in-cell, Vlasov/Boltzmann, and hybrid codes in 1, 2, or 3 dimensions. 

We posit that one should be able to calculate the key phase-space-based quantities in well-resolved satellite observations, although the practical use of KPS and KSR requires careful consideration of not only instrumental cadence and sampling, but also the number of spacecraft and their configuration. KPS requires well-resolved 3-D phase-space densities with sufficient counts in velocity space, as well as measurements of the local spatial derivatives of the bulk flow velocity. KSR additionally requires the ability to estimate local spatial derivatives of the phase-space density (normalized to the local number density). For a single spacecraft, this requires employing the Taylor-hypothesis for time-to-space conversion. MMS is specifically well-suited to measure KPS and KSR due to its measurement cadence and the tetrahedron flying configuration. For cases where continuous and smooth phase-space density measurements are limited, leading to data gaps, reconstruction methods using custom-designed basis functions, such as Slepian functions \citep{Das&Terres_2025_ApJ,Das&Terres_2025_ApJ_b} could be useful in retrieving lost phase space data. 
The applicability of MMS depends on the availability of intervals with all four Dual Electron Spectrometers of the Fast Plasma Investigation suite \citep{Pollock_FPI_2016} to use the tetrahedral gradient estimation, \textit{i.e.,} the curlometer \citep{Harvey1998, Dunlop_2002_JGR} technique, to compute strain-rate tensor elements \citep{Chasapis18}. (However, see Ref.~\citep{singh2026integration}, where it is suggested that even when one spectrometer is unavailable, the strain-rate tensor could be calculable in many three-spacecraft intervals.) 

NASA's forthcoming HelioSwarm mission \citep{klein_helioswarm_2023} is unlikely to be used to measure the kinetic quantities employed in the present study.  It is designed to deliver nine-point measurements of magnetic field, plasma density, and flow fluctuations across MHD-to-sub-ion scales, together with full proton distribution-function measurements at the Hub spacecraft. It may help connect multiscale field/flow structure to ion heating signatures and to proton velocity-space structure. Because the Nodes will provide only radial ion distribution information and derived plasma moments, the ion pressure-strain interaction will not be directly measurable. 
Likewise, electron pressure-strain interaction will not be measurable because electron VDF measurements are likely to only be measured at the Hub. 
However, the quantities in the present study may be measurable in the ESA mission concept Plasma Observatory \citep{Retino_PlasmaObservatory_2022}. It would deploy seven spacecraft in a multiscale configuration designed to sample Earth's magnetospheric system simultaneously at fluid and ion scales. A mothercraft would provide higher-time-resolution field and particle measurements down to sub-ion scales. Plasma Observatory would provide electron and ion VDF measurements at the daughtercraft, which would allow for the measurement of the pressure tensor and the strain-rate tensor.

We note several avenues that warrant further exploration.
This study relies on a single high-resolution magnetic reconnection simulation, which uses an antiparallel 2.5-D geometry, and a single snapshot in the growth phase of reconnection where the simulation domain does not allow for a steady state. Future work should explore simulations where fully ion-recoupled reconnection, \textit{i.e.,} a quasi-steady state, is achievable, as it is known that the pressure-strain interaction changes between the onset and steady-state phases of reconnection \citep{Barbhuiya_phases_2025}. 
It would also be useful to change the reconnection simulation parameters, \textit{i.e.,} by adding a guide field and/or changing upstream plasma $\beta$, and repeat the present study parametrically. 

Other important future work is to apply the approach of this study to turbulent and wave-rich environments. 
An additional parameter dimension of interest is scale separation (\textit{e.g.,} the electron-ion mass ratio), which can influence electron-scale current sheet structure and the development and amplitude of drift- and shear-driven waves in and around thin current layers \citep{Dubois2026_JGR}. Applying the present phase-space framework to simulations with larger ion-to-electron mass ratio and/or fully 3D reconnection \citep{Greess2021_JGR}, with explicit attention to shear-driven lower-hybrid activity in compressed layers \citep{dubois2022PRL}, could help determine how wave-mediated anomalous effects and directly shear-driven energization compete in thin current sheets.

The three locations we analyze in the present study have populations that are identifiable by eye but are not as well isolated in velocity space as, {\it e.g.,} reflected and beam ion populations found in a recent study of collisionless shocks \citep{Juno_2023_ApJ}. Thus, in the present work, we perform a mostly qualitative analysis of the relative importance of the different populations.  
This choice is intentional because the relevant electron populations overlap in velocity space, even when they are visually identifiable as coherent structures in the reduced VDFs and phase-space-based quantities. Thus, assigning sharp masks to individual populations would require additional assumptions and could make population-integrated values depend on the chosen demarcation procedure. We therefore use the reduced phase-space structures to identify the dominant physical contributions to KPS and KSR, qualitatively, while deferring objective population separation and population-resolved integrations to future work.
To perform a more quantitative analysis, an approach that separates the contributions due to different populations would be desirable \citep{Goldman_PoP_2021}.
Machine learning (ML) may provide a path forward to demarcate different populations within a VDF \citep{Sano2025analysis}.
Such an ML tool could also prove useful in low-cadence measurements of phase-space densities by spacecraft where data gaps exist. With careful training, uncertainty quantification including instrument effects, and adaptation to spacecraft data, it could be possible to automate the identification of the particle populations dominantly responsible for normal flow vs.~sheared flow-driven energization heating in large MMS and other \textit{in-situ} spacecraft datasets. This could be invaluable in improving the interpretation of strongly structured non-Maxwellian phase-space densities.

Another key area for future work is to go beyond the alternative kinetic pressure-strain to include other terms that can change the phase space internal energy density.
As shown in Eq.~\ref{EQ:PSINTEQ2}, seven terms in total can contribute to the time evolution of $w_{{\rm int}}$.
Each term represents a different mechanism that can evolve $w_{\rm int}$, out of which only one was studied here and in Ref.~\citep{Conley_2024_PoP}. 
The key point is that the pressure-strain interaction and the kinetic pressure-strain describe only one mechanism for the evolution of the internal energy density, and a complete kinetic picture of internal energy evolution requires understanding the roles of the other terms.


\begin{acknowledgments}
We gratefully acknowledge support from NASA Grants 80NSSC24K0172 (PAC), 80NSSC19M0146 (PAC), 80NSSC24K0552 (GGH), and 80NSSC23K0099 (JMT), NSF Grant PHY‐1804428 (PAC), DOE Grant DE‐SC0020294 (PAC), and Royal Society University Research Fellowship awards URF\textbackslash R\textbackslash 251029 and URF\textbackslash R1\textbackslash 201286 (JES). This research was partially supported by the International Space Science Institute (ISSI) in Bern, through ISSI International Team Project 23‐588 (“Unveiling Energy Conversion and Dissipation in Non‐Equilibrium Space Plasmas”). This research used resources of the National Energy Research Scientific Computing Center (NERSC), a U.S. Department of Energy Office of Science User Facility located at Lawrence Berkeley National Laboratory, operated under Contract DE‐AC02‐05CH11231 using NERSC Award FES‐ERCAP0027083.
\end{acknowledgments}

\section*{Data Availability Statement}
The data supporting the findings of this study are openly available in Zenodo at 
https://doi.org/10.5281/zenodo.20453508

\appendix

\section{Velocity-space integral of the force term in Eq.~\ref{EQ:PSINTEQ1}} \label{appensec:A}

\begin{figure*}[t]
    \centering
    \includegraphics[width=\textwidth]{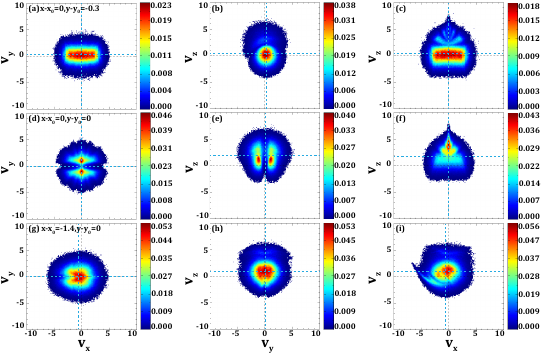}
    \caption{Reduced electron VDFs at three locations, with dark sky blue dashed lines denoting the bulk flow velocity components, $u_{ex}$, $u_{ey}$, and $u_{ez}$. (a)-(c) $(x-x_0,y-y_0)=(0,-0.3)$, near the upstream edge of the EDR, with $(u_{ex},u_{ey},u_{ez})=(0.00039,0.23,0.51)$. (d)-(f) $(x-x_0,y-y_0)=(0,0)$, surrounding the X-line, with $(u_{ex},u_{ey},u_{ez})=(-0.00026,0.0059,1.9)$. (g)-(i) $(x-x_0,y-y_0)=(-1.4,0)$, near the outflow edge of the EDR, with $(u_{ex},u_{ey},u_{ez})=(-0.64,-0.0015,0.95)$. The left, middle, and right columns display VDFs reduced into the $v_x-v_y$, $v_y-v_z$, and $v_x-v_z$ planes, respectively.}
   \label{fig:rVDFcombined}
\end{figure*}

Rewriting the velocity-space integral of the force term in Eq.~\ref{EQ:PSINTEQ1} in Einstein summation notation, we obtain
\begin{equation}
    \int d^3v \frac{1}{2} v^{\prime 2} \nabla_v \cdot \left({\bf F} f \right) = \int d^3 v \frac{ v^{\prime 2}}{2} \frac{\partial}{\partial v_k} (f F_k). \nonumber
\end{equation}
Integration by parts gives
\begin{equation}
 \int d^3v \frac{1}{2} v^{\prime 2} \nabla_v \cdot \left({\bf F} f \right) = \int d^3v \frac{\partial}{\partial v_k} \left(\frac{ v^{\prime 2}}{2} f F_k\right) - \int d^3v f F_k \frac{\partial}{\partial v_k} \left(\frac{ v^{\prime 2}}{2}\right)
= \int \frac{ v^{\prime 2}}{2} f {\bf F} \cdot d{\bf S}
- \int v_k^{\prime} f F_k d^3v = - \int d^3v {\bf v}^\prime \cdot {\bf F} f, \nonumber
\end{equation}
where $d{\bf S}$ is the surface at infinity in velocity space and $\int (v^{\prime 2}/2) f {\bf F} \cdot d{\bf S}$ is 
zero for a normalizable $f$. For the remaining term, we use the electromagnetic force, 
\begin{equation}
    \int d^3v {\bf v}^\prime \cdot {\bf F} f = \int d^3v {\bf v}^\prime \cdot \left( q {\bf E} + {\bf v} \times \frac{{\bf B}}{c} \right) f. \nonumber
\end{equation}
As the electric field term on the right-hand side is linear in
${\bf v}^\prime$, it integrates to zero. Next, we write ${\bf v}^\prime = {\bf v} - {\bf u}$ inside the integral on the right-hand side, which gives
\begin{equation}
    \int d^3v {\bf v}^\prime \cdot \left({\bf v} \times \frac{{\bf B}}{c} \right) f  = \int d^3v {\bf v} \cdot \left({\bf v} \times \frac{{\bf B}}{c} \right) f - \int d^3v {\bf u} \cdot \left({\bf v} \times \frac{{\bf B}}{c} \right) f \nonumber
\end{equation}
The first term is zero using a vector identity, and the second term gives 
\begin{equation}
    \int d^3v {\bf u} \cdot \left({\bf v} \times \frac{{\bf B}}{c} \right) f  = {\bf u} \cdot  \int d^3v  \left({\bf v} \times \frac{{\bf B}}{c} \right) f = {\bf u} \cdot \left({\bf u} \times \frac{{\bf B}}{c} \right). \nonumber
\end{equation}
which is also zero. Thus, the velocity space integral of the force term in Eq.~\ref{EQ:PSINTEQ1} vanishes.

\section{Decomposition of the alternate kinetic pressure-strain in Cartesian coordinates}\label{appensec:B}

\textcolor{black}{In Ref.~\citep{Conley_2024_PoP}, $\widetilde{K}_{\rm PS}$ is decomposed into $\widetilde{K}_{\mathcal{P} \theta}$ and $\widetilde{K}_{\rm PiD}$, thereby focusing on the phase-space analogs of the fluid quantities 
pressure dilatation and ${\rm Pi-D}$ that capture compressible and incompressible physics, respectively \citep{Yang17,yang_PRE_2017}.} 
Here, we provide the expressions for the two phase-space analogs in Cartesian coordinates to gather the two decompositions of $\widetilde{K}_{\rm PS}$ in one place. 
Adding and subtracting a common term in Eq.~\ref{eq:altKPS_decomp} yields
\begin{equation}
\widetilde{K}_{\rm PS}  = -\sum_j \frac{1}{3}m v_j^{\prime 2} f (\boldsymbol{\nabla} \cdot {\bf u}) \\
- \Bigl(\sum_{j,k} m v_j^{\prime} v_k^{\prime} f \partial_j u_k -  \sum_j \frac{1}{3}m v_j^{\prime 2} f (\boldsymbol{\nabla} \cdot {\bf u}) \Bigr), \label{eq:altKPS_decomp2}
\end{equation}
where the first and second terms (including the negative sign) are $\widetilde{K}_{\mathcal{P} \theta}$ and $\widetilde{K}_{\rm PiD}$, respectively. For a 3-D system, 
$\widetilde{K}_{\mathcal{P} \theta}$ and $\widetilde{K}_{\rm PiD}$ are written explicitly as 
\begin{equation}
\widetilde{K}_{\mathcal{P} \theta} = -\frac{1}{3} \Bigl(m v_x^{\prime 2}  
+ m v_y^{\prime 2}  
+ m v_z^{\prime 2} \Bigr) f \boldsymbol{\nabla} \cdot {\bf u} \\
 \label{eq:Kptheta}
\end{equation}
and
\begin{equation}
\begin{split}
\widetilde{K}_{\rm PiD} = \frac{1}{3} \Bigl(m v_x^{\prime 2} + m v_y^{\prime 2}  + m v_z^{\prime 2} \Bigr) 
f \boldsymbol{\nabla} \cdot {\bf u}-f\Bigl[m v_x^{\prime 2} \partial_x u_x + m v_y^{\prime 2} \partial_y u_y + m v_z^{\prime 2} \partial_z u_z 
&+ m v_x^{\prime} v_y^{\prime} \left(\partial_x u_y + \partial_y u_x\right) \\
&+ m v_y^{\prime} v_z^{\prime} \left(\partial_y u_z + \partial_z u_y\right) \\
&+ m v_x^{\prime} v_z^{\prime} \left(\partial_x u_z + \partial_z u_x\right) \Bigr]. \label{eq:KPiD}
\end{split}
\end{equation}

\section{Reduced VDFs at the locations of analysis and their associated pressure tensor elements} \label{appensec:C}
 
For completeness, Fig.~\ref{fig:rVDFcombined} displays reduced VDFs at the three locations discussed in the main body of the paper. The left, center, and right columns are reduced in $v_z$, $v_x$, and $v_y$, respectively.
Panels (a)-(c) are at $(x-x_0,y-y_0)=(0,-0.3)$, panels (d)-(f) are at $(x-x_0,y-y_0)=(0,0)$, and panels (g)-(i) are at $(x-x_0,y-y_0)=(-1.4,0)$.
The dark sky blue dashed lines denote the bulk flow velocity components. All three electron VDFs are distinctly non-Maxwellian.

To support the treatment in Sec.~\ref{sec:results}, we now describe the phase-space origin of the electron pressure tensor elements at the locations of analyses in this study.
We demarcate the bulk flow velocities using the dark blue sky dashed lines in each panel.  Because the pressure-tensor elements are dependent only on the peculiar/random motion of the particles, the asymmetries of the VDF that lead to non-zero pressure tensor elements are determined relative to the bulk flow, {\it i.e.,} in terms of $\mathbf{v}^\prime$. 

For the position at $(x-x_0,y-y_0) = (0,-0.3),$ there is a symmetry in the $x$ direction. Thus, the electron VDF is symmetric in $v'_x$, which implies that $P_{xy} \approx P_{xz} \approx 0$.
In the $v'_y-v'_z$ plane, the reduced VDF is skewed in the second and fourth quadrants, implying $P_{yz}<0$. As the VDF is more extended in $v'_x$ than in $v'_y$ \citep{Egedal13}, $P_{xx} > P_{yy}$. The spread of the VDF in $v'_z$ due to the striation electron population leads to $P_{zz}> P_{xx}$.

At $(x-x_0,y-y_0) = (0,0)$, the symmetry in $x$ and $y$ leads to symmetry in the VDF in $v'_x$ and $v'_y$, resulting in $P_{xy} \approx P_{yz} \approx P_{xz}\approx 0$. 
Since the VDF is more elongated in $v'_x$ than in $v'_y$, $P_{xx}> P_{yy}$. The presence of the striation electron population elongates the VDF in $v'_z$, leading to $P_{zz}>P_{xx}$.
 
At $(x-x_0,y-y_0)=(-1.4,0)$, there is symmetry in the $y$ direction and therefore the VDF is symmetric in $v'_y$, implying that $P_{xy} \approx P_{yz} \approx 0$. 
In the $v'_x-v'_z$ plane, the reduced VDF is skewed in the first and third quadrants, so $P_{xz} > 0$. 
The VDF is elongated more in $v'_x$ than $v'_y$, resulting in $P_{xx}>P_{yy}$. Due to the rotated striation population, there is an effective elongation in $v'_z$ that leads to $P_{zz} > P_{xx}$. 

\bibliography{Strain_rate_Particle_Correlation}{}
\bibliographystyle{aasjournalv7}

\end{document}